\documentclass[ aps,%
floatfix,%
final,%
notitlepage,%
oneside,%
onecolumn,%
nobibnotes,%
nofootinbib,%
superscriptaddress,%
showpacs,%
]%
{revtex4-1}

\usepackage{graphicx}

\usepackage{amsmath}
\usepackage{hyperref}
\usepackage{booktabs}

\usepackage[english]{babel}

\newcommand{\Q}{\mathcal{Q}}

\newcommand{\M}{\mathcal{M}}
\newcommand{\JP}{{J/\psi}}
\newcommand{\fL}[1]{f_{#1}}
\newcommand{\phiL}[1]{\phi_{#1}}
\newcommand{\NRQCD}{\mathrm{NRQCD}}
\newcommand{\fN}[1]{F_{#1}}
\newcommand{\GeV}{\,\mathrm{GeV}}
\newcommand{\QCD}{\mathrm{QCD}}
\newcommand{\Br}{\mathrm{Br}}
\allowdisplaybreaks

\begin{document}
\title{Double Charmonia Production in Exclusive $Z$ Boson Decays}

\author{A.~K.~Likhoded}
\email{Anatolii.Likhoded@ihep.ru@ihep.ru}
\affiliation{``Institute for High Energy Physics" NRC ``Kurchatov Institute'', 142281, Protvino, Russia}

\author{A.~V.~Luchinsky}
\email{alexey.luchinsky@ihep.ru}
\affiliation{``Institute for High Energy Physics" NRC ``Kurchatov Institute'', 142281, Protvino, Russia}

\begin{abstract}
This article is devoted to systematic analysis of double charmonium production in exclusive $Z$ boson decays in the framework of Nonrelativistic Quantum Chromodynamic (NRQCD) and leading twist light-cone (LC) models. Theoretical predictions for branching fractions of all considered decays are presented. According to obtained results in the case of allowed by helicity suppression rule processes the effect of internal quark motion increases the branching fraction by a factor 1.5, while for forbidden reactions LC predictions are strictly zero, while NRQCD ones are significantly smaller than for allowed.
\end{abstract}
\maketitle

\section{Introduction}
\label{sec:introduction}

Interplay of long-distance and short-distance effects in Quantum Chromodynamic (QCD) is a long-standing puzzle in the physics of elementary particles. Due to asymptotic freedom at short distances strong coupling constant $\alpha_s$ is small, so the perturbation theory can be used to describe this region. At  long distances, on the other hand, $\alpha_s\sim 1$ and some nonpeturbative methods should be used.

In the field of heavy quarkonia production and decays one of such methods is the Nonrelativistic Quantum Chromodynamic (NRQCD) \cite{Bodwin:1994jh}, that exploits the fact, that heavy quark's mass $m_Q$ is large in comparison with  typical scale of the strong interaction $\Lambda_\mathrm{QCD}$. As a result quark's internal velocity inside the meson $v$ should be expected to be small and the amplitude of the process can be written as a series over this parameter. At the leading order (LO) of such expansion internal motion of the quarks is neglected completely and the coefficient, that describe the long-distance part of the interaction, can be obtained, for example, using different potential models or comparison with the experiment. It should be noted, however, that in the case of charmonia production this expansion parameter is actually not small ($v^2\sim\alpha_c(m_c)\sim 0.3$) so there could be significant errors at the leading order. As an example one can name double charmonia production in $e^+e^-$ annihilation at $B$-factories Belle and BaBar \cite{Abe:2004ww,Aubert:2005tj}, when LO  NRQCD predictions turned out to be about an order of magnitude smaller that the experimental results.

An alternative method for studying charmonia production at high energy interactions is the formalism of amplitude expansion on the light cone (LC) \cite{Chernyak:1983ej}. In this method the internal motion of the quark is described by nonpertubative distribution amplitudes and the small expansion parameter is a chirality factor $m_q/E$, where $m_q$ and $E$ are mass of the produced quark and typical energy of the reaction respectively. Such approach, obviously, is best suitable for description of light meson ($\pi$, $\rho$, etc.) production, but it can be used also for charmonium meson production at high energies. For example, mentioned above disagreement in double charmonia production at $B$-factories was explained in the LC framework \cite{Bondar:2004sv,Braguta:2005kr}. The other example is double charmonium production in exclusive bottomonia decays \cite{Braguta:2009df, Braguta:2009xu, Chen:2012ih}. Single charmonium production in radiative $Z$-boson decays was also studied both in NRQCD and LC \cite{Guberina:1980dc, Grossmann:2015lea, Alte:2017ycm, Luchinsky:2017jab}

In the present paper a systematic analysis of double charmonia production in exclusive $Z$-boson decays is presented.
In the next section we show analytical results obtained using both NRQCD and LC approaches. Helicity suppression rules are also discussed in details. 
Numerical predictions for the branching fractions are shown in section \ref{sec:numerical-results} and the last section is reserved for conclusion.

\section{Analytical Results}
\label{sec:analytical-results}

\subsection{NRQCD}
\label{sec:nrqcd}

One of the models that can be used to describe the processes under consideration is Nonrelativistic Quantum Chromodynamics (NRQCD)\cite{Bodwin:1994jh}. This model uses the fact that heavy quark mass $m_c$ is large in comparison with typical QCD scale $\Lambda_\QCD$. As a result, the strong coupling constant $\alpha_s(m_c)$ and typical squared internal quarks' velocity $v^2_c\sim\alpha_s(m_c)$ can be considered to be small and the matrix element of the reaction is written as a series over these parameters. In more details, the hard part of the amplitude, that describe production of heavy quark-antiquark pair is calculated using perturbative QCD, while hadronization of this pair into the final meson is described by NRQCD matrix elements. It is clear that the latter part of the process is essentially nonperturbative, so some other methods should be used to obtain the numerical values of these matrix elements. In the framework of NRQCD one should take into account contributions of both color-singlet (CS) and color-octet (CO) components of the heavy quarkonium Fock state. Color-singlet matrix elements can be determined, for example, using experimental values of the widths of corresponding decays (e.g. $\JP\to\mu\mu$ or $\chi_{c0,2}\to\gamma\gamma$). The definition of CO matrix elements is more involved and we will neglect color-octet contributions in the following.

 \begin{figure}
   \centering
   \includegraphics[width=0.9\textwidth]{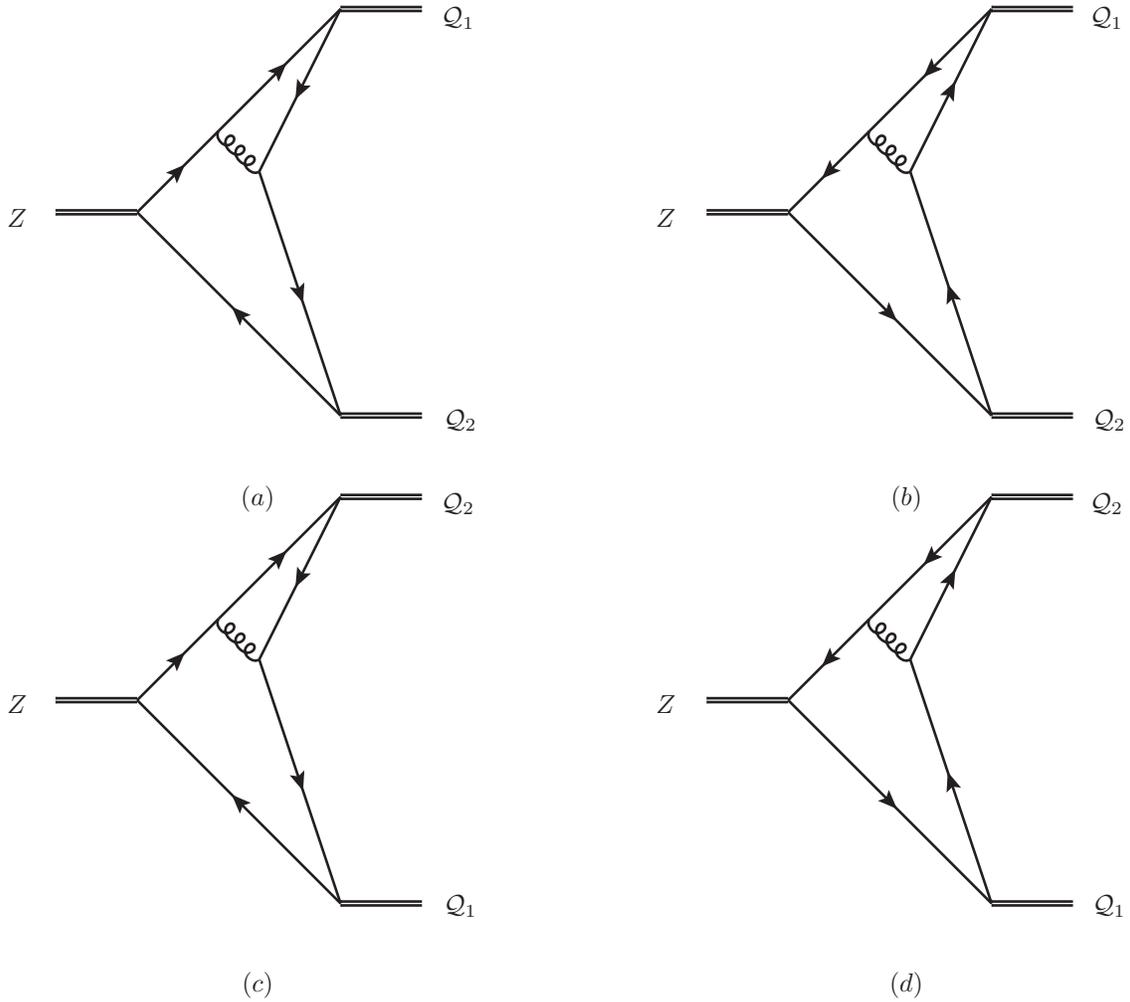}
   \caption{Feynman diagrams for $Z\to\Q_1\Q_2$ process}
   \label{fig:diags}
 \end{figure}

Leading order Feynman diagrams that describe double charmonia production in exclusive $Z$ boson decays $Z\to\Q_1\Q_2$ are shown in fig.~\ref{fig:diags}. Using the projection technique, described, for example, in \cite{Braaten:2002fi} it is easy to obtain analytical expressions for the amplitudes of the considered processes. It is convenient to define mesons' constants in the following way:
\begin{align}
  \label{eq:f_NRQCD_S}
 \fN{\eta_c} &= \fN{\JP} = \sqrt{\frac{\langle O_1\rangle_{\JP}}{m_c}} \\
  \label{eq:f_NRQCD_P}
 \fN{h_c} &= \sqrt{3} \fN{\chi_{c0}} = \frac{1}{\sqrt{2}}\fN{\chi_{c1}} = \sqrt{\frac{3}{2}} \fN{\chi_{c2}} = \sqrt{\frac{\langle O_1\rangle_{h_C}}{m_c^3}},   
\end{align}
where $\langle O_1\rangle_{J/\psi,h_c}$ are defined in \cite{Braaten:2002fi} NRQCD matrix elements for $S$- and $P$-wave charmonium mesons.
If we define the helicities of initial bozon and final charmonia as $\lambda_Z$ and $\lambda_{1,2}$ respectively, the squared matrix element of $Z\to\Q_1\Q_2$ reaction can be written in the form
\begin{align}
  \label{eq:NRQCD_sqr}
  \left|\M(Z(\lambda_Z)\to\Q_1(\lambda_1)\Q_2(\lambda_2)  \right|^2 &=
   \frac{4096\pi^2\alpha_s^2}{81}
    c_{V,A}^2
    \frac{\fN{\Q_1}^2\fN{\Q_2}^2}{M_Z^2} C^{\Q_1\Q_2}_{\lambda_1\lambda_2},
\end{align}
where $c_{V,A}$ are vector and axial coupling constants of $Zc\bar{c}$ vertex
\begin{align}
  \label{eq:Zvert}
  c_V \gamma_\mu + c_A \gamma_\mu\gamma_5,
\end{align}
and $C^{\Q_1\Q_2}_{\lambda_1\lambda_2}$ coefficients are given in the appendix \ref{sec:nrqcd-matr-elem}. 
Using these relations it is easy to calculate total widths of the considered decays. In is clear, that the processes $Z\to\eta_c\eta_c$ and $Z\to\chi_{c0}\chi_{c0}$ are strictly forbidden by Boze symmetry. As for all other processes, they can be divided into two groups. The widths of the first group of processes
\begin{align}
  \label{eq:NRQCD_widths_allowed1}
\Gamma_\NRQCD(Z\to h_c \JP) &=\frac{256 \pi  \alpha_s^2 \beta ^3 c_A^2 \fN{h_c}^2 \fN{\JP}^2 \left(4608 r^6-1504 r^4+128 r^2+1\right)}{243 M_Z^3}\\ 
\Gamma_\NRQCD(Z\to h_c \eta_c) &=\frac{256 \pi  \alpha_s^2 \beta  c_V^2 \fN{h_c}^2 \fN{\eta_c}^2 \left(36864 r^8-8704 r^6+832 r^4-40 r^2+1\right)}{243 M_Z^3}\\ 
\Gamma_\NRQCD(Z\to h_c \chi_{c1}) &=\frac{256 \pi  \alpha_s^2 \beta ^3 c_V^2 \fN{h_c}^2 \fN{\chi_{c1}}^2 \left(13312 r^6-3328 r^4+208 r^2+1\right)}{243 M_Z^3}\\ 
\Gamma_\NRQCD(Z\to \JP \chi_{c0}) &=\frac{256 \pi  \alpha_s^2 \beta  c_V^2 \fN{\JP}^2 \fN{\chi_{c0}}^2 \left(36864 r^8+9728 r^6-6848 r^4+728 r^2+1\right)}{243 M_Z^3}\\ 
\Gamma_\NRQCD(Z\to \JP \chi_{c2}) &=\frac{256 \pi  \alpha_s^2 \beta  c_V^2 \fN{\JP}^2 \fN{\chi_{c2}}^2 \left(92160 r^8+19712 r^6-3008 r^4+80 r^2+1\right)}{243 M_Z^3}\\ 
\Gamma_\NRQCD(Z\to \eta_c \chi_{c0}) &=\frac{256 \pi  \alpha_s^2 \beta ^3 c_A^2 \fN{\eta_c}^2 \fN{\chi_{c0}}^2 \left(1-8 r^2\right)^2}{243 M_Z^3}\\ 
\Gamma_\NRQCD(Z\to \eta_c \chi_{c2}) &=\frac{256 \pi  \alpha_s^2 \beta ^3 c_A^2 \fN{\eta_c}^2 \fN{\chi_{c2}}^2 \left(3456 r^6-1136 r^4+104 r^2+1\right)}{243 M_Z^3}\\ 
\Gamma_\NRQCD(Z\to \chi_{c0} \chi_{c1}) &=\frac{256 \pi  \alpha_s^2 \beta  c_A^2 \fN{\chi_{c0}}^2 \fN{\chi_{c1}}^2 \left(524288 r^{10}-454656 r^8+138752 r^6-17648 r^4+840 r^2+1\right)}{243 M_Z^3}\\ 
\label{eq:NRQCD_widths_allowed2}
\Gamma_\NRQCD(Z\to \chi_{c1} \chi_{c2}) &=\frac{256 \pi  \alpha_s^2 \beta  c_A^2 \fN{\chi_{c1}}^2 \fN{\chi_{c2}}^2 \left(1064960 r^{10}-466944 r^8+78656 r^6-5024 r^4+96 r^2+1\right)}{243 M_Z^3}
%
\end{align}
are not suppressed by the chirality factor
\begin{align}
  \label{eq:r}
  r &= \frac{m_c}{M_Z} \approx 0.02.
\end{align}
The widths of all other processes
\begin{align}
  \label{eq:NQRCD_widths_suppressed1}
\Gamma_\NRQCD(Z\to \eta_c \JP) &=\frac{8192 \pi  \alpha_s^2 \beta ^3 c_V^2 \fN{\eta_c}^2 \fN{\JP}^2 r^2}{243 M_Z^3}\\ 
\Gamma_\NRQCD(Z\to \eta_c \chi_{c1}) &=\frac{2048 \pi  \alpha_s^2 \beta ^3 c_A^2 \fN{\eta_c}^2 \fN{\chi_{c1}}^2 r^2 \left(1-4 r^2\right)^2}{27 M_Z^3}\\ 
\Gamma_\NRQCD(Z\to \JP \JP) &=\frac{1024 \pi  \alpha_s^2 \beta ^5 c_A^2 \fN{\JP}^4 r^2}{243 M_Z^3}\\ 
\Gamma_\NRQCD(Z\to \JP \chi_{c1}) &=\frac{8192 \pi  \alpha_s^2 \beta  c_V^2 \fN{\JP}^2 \fN{\chi_{c1}}^2 r^2 \left(576 r^6+104 r^4-24 r^2+1\right)}{243 M_Z^3}\\ 
\Gamma_\NRQCD(Z\to \chi_{c0} \chi_{c2}) &=\frac{2048 \pi  \alpha_s^2 \beta ^5 c_A^2 \fN{\chi_{c0}}^2 \fN{\chi_{c2}}^2 r^2 \left(7-24 r^2\right)^2}{81 M_Z^3}\\ 
\Gamma_\NRQCD(Z\to \chi_{c0} h_c) &=\frac{8192 \pi  \alpha_s^2 \beta ^3 c_V^2 \fN{\chi_{c0}}^2 \fN{h_c}^2 r^2 \left(3-8 r^2\right)^2}{243 M_Z^3}\\ 
\Gamma_\NRQCD(Z\to \chi_{c1} \chi_{c1}) &=\frac{1024 \pi  \alpha_s^2 \beta  c_A^2 \fN{\chi_{c1}}^4 r^2 \left(32 r^4-18 r^2+1\right)^2}{243 M_Z^3}\\ 
\Gamma_\NRQCD(Z\to \chi_{c2} \chi_{c2}) &=\frac{1024 \pi  \alpha_s^2 \beta ^5 c_A^2 \fN{\chi_{c2}}^4 r^2 \left(348 r^4-36 r^2+1\right)}{81 M_Z^3}\\ 
\Gamma_\NRQCD(Z\to \chi_{c2} h_c) &=\frac{8192 \pi  \alpha_s^2 \beta ^3 c_V^2 \fN{\chi_{c2}}^2 \fN{h_c}^2 r^2 \left(1312 r^4-120 r^2+3\right)}{243 M_Z^3}\\ 
\Gamma_\NRQCD(Z\to h_c h_c) &=\frac{1024 \pi  \alpha_s^2 \beta ^5 c_A^2 \fN{h_c}^4 r^2}{243 M_Z^3}
  \label{eq:NQRCD_widths_suppressed2}
\end{align}
on the other hand, are zero in $r\to 0$ limit.

\subsection{Selection Rules}
\label{sec:selection-rules}

The effect of chirality suppression can be explained using surprisingly simple arguments. It is well known, that for high energy strong interactions some selection rules, e.g. helicity conservation, should be satisfied. This rule is caused by the fact that, due to vector nature of quark-gluon vertex in massless quark limit, this interaction conserves quark's helicity. This property is true also in the case of $Z$-boson-quark interaction, so the same rule should be satisfied also in the case of considered reactions. Since there are only two hadrons in these reactions, we should have
\begin{align}
  \lambda_1 + \lambda_2 &= 0,
\end{align}
where $\lambda_{1,2}$ are the helicities of final charmonia. From orbital momentum conservation, on the other hand, it follows that
\begin{align}
  -1 &\le \lambda_1-\lambda_2=\lambda_Z \le 1,
\end{align}
where $\lambda_Z$ is the projection of $Z$-boson spin on the final charmonium momentum direction.
It is easy to check that only $\lambda_1=\lambda_2=0$ configuration satisfies these restrictions, so only longitudinally polarized mesons can be produced.
Such configuration, however, is  allowed not for all processes. In the leading twist approximation the polarization vector of longitudionally polarized particle is proportional to its momentum. If total amplitude of the process is proportional to absolutely antisymmetric tensor $e_{\alpha\beta\gamma\delta}$, it is actually equal to zero, since we do not have enough independent vectors. In order to quatitize this rule it is convenient to introduce so called ``naturality'' quantum number of the meson, defined as
\begin{align}
  \label{eq:narurality}
  \sigma &= P(-1)^J,
\end{align}
where $P$ and $J$ are space parity and total spin of the particle. For mesons with negative naturality the quark-meson projector will contain absolutely anti-symmetric tensor. It is clear, on the other hand, in the case of positive charged parity of the final state an axial component of $Z$ boson will be contributing and $\gamma_5$ Dirac matrix is present in the corresponding interaction vertex. As a result, allowed in the leading twist approximation processes should satisfy the following selection rule:
\begin{align}
  C_1 C_2 \sigma_1 \sigma_2 &= -1,
\end{align}
where $C_{1,2}$ and $\sigma_{1,2}$ are charge parities and naturalities of final charmonium mesons. The same rule can be formulated as naturality conservation rule if one, according to definition \eqref{eq:narurality}, assign the values $\sigma_Z=1$ and $-1$ to vector and axial components of $Z$-boson respectively.
It is easy to check, that only the following processes satisfy listed above restrictions:
\begin{align}
  \label{eq:allowed_proc}
  Z \to \eta_c \chi_{c0}, \eta_c \chi_{c2}, \eta_c h_c, 
           \JP \chi_{c0}, \JP \chi_{c2}, \JP h_c,
          \chi_{c1} \chi_{c0}, \chi_{c1}\chi_{c2}, \chi_{c1} h_c
  = (\eta_c, \JP,\chi_{c1}) \otimes (\chi_{c0}, \chi_{c2}, h_c), 
\end{align}
i.e. processes listed in eq.~\eqref{eq:NRQCD_widths_allowed1} -- \eqref{eq:NRQCD_widths_allowed2}. It is
interesting to note, that in listed processes one of the produced
mesons is $\eta_c$, $\JP$ or $\chi_{c1}$, while the other should be
either $\chi_{c0,2}$ or $h_c$. Production of all other states
(e.g. $\JP\JP$) is suppressed by mentioned above chirality suppression
factor $r$. This fact agrees perfectly with presented in eq.~\eqref{eq:NQRCD_widths_suppressed1} -- \eqref{eq:NQRCD_widths_suppressed2} results.


\subsection{Light Cone Expansion}
\label{sec:light-cone-expansion}

An alternative way to consider listed above processes is light-cone expansion formalism \cite{Chernyak:1983ej}. In this approach the amplitude of the  reaction is written as a series over a small light quark mass $m_q$. It is clear, that this model is most suitable for description of light meson production in high-energy experiments. Recently, however, an attempts were made to use it also for heavy quarkonia production at B-factories Belle and BaBar at $\sqrt{s_{ee}}=10.6$ GeV (see \cite{Abe:2004ww, Bondar:2004sv, Aubert:2005tj, Braguta:2005kr}) and agreement between obtained theoretical predictions and experimental data shows us that light-cone formalism can be used also to describe reactions with heavy quarkonium mesons. Since in $Z$-boson decays the characteristic energy $M_Z\gg\sqrt{s_{ee}}$, one can safely use this model for analysis of charmonium production in these reactions.
The amplitude of the considered process can be written as a convolution
\begin{align}
  \label{eq:LC_main}
  \M(Z\to\Q_1\Q_2) &\sim \fL{\Q_1}\fL{\Q_2} \int_{-1}^1 d\xi_1 d\xi_2
                     \phiL{\Q_1}(\xi_1)  \phiL{\Q_2}(\xi_2) H(\xi_1,\xi_2),
\end{align}
where $\xi_{1,2}=x_{1,2}-\bar{x}_{1,2}$ are the differences between
momentum fractions carried by quark and antiquark respectively,  $H(\xi_1,\xi_2)$ is a short-distance part of the amplitude, that can be calculated perturbativelly, while leptonic constants $\fL{\Q_{1,2}}$ and distribution amplitudes (DAs) $\phiL{\Q_{1,2}}(\xi_{1,2})$ describe long-distance effects and should be studied using non-perturbative methods such as QCD sum rules.

The distribution amplitudes that enter in \eqref{eq:LC_main} are defined as
\begin{align}
  \langle\Q_L(p)
  \left|\bar{c}^i_\alpha(z)[z,-z]c^j_\beta(-z)\right|0\rangle
  &=
    \left(\hat{p}\right)_{\alpha\beta}\frac{\fL{\Q}}{4}
    \frac{\delta^{ij}}{3}
    \int\limits_{-1}^1 \phiL{\Q}(\xi)d\xi
\end{align}
for $\sigma$-even states $\Q=\JP$, $\chi_{0,2}$ and
\begin{align}
  \label{eq:DA_odd}
  \langle\Q_L(p)
  \left|\bar{c}^i_\alpha(z)[z,-z]c^j_\beta(-z)\right|0\rangle
  &=
    \left(\hat{p}\gamma_5\right)_{\alpha\beta}\frac{\fL{\Q}}{4}
    \frac{\delta^{ij}}{3}
    \int\limits_{-1}^1 \phiL{\Q}(\xi)d\xi
\end{align}
for $\sigma$-odd states $\Q=\eta_c$, $\chi_{c1}$, and $h_c$. In the above expressions $\alpha,\beta$ and $i,j$ are spinor and colour indices of quark and antiquark respectively. The normalization condition for the distribution amplitudes is
\begin{align}
  \label{eq:norm_xi_even}
  \int_{-1}^1\phiL{\Q}(\xi)d\xi &=1,\qquad   \int_{-1}^1 \xi\phiL{\Q}(\xi)d\xi=1 
\end{align}
for $\xi$-even  ($\Q=\eta_c,\JP, \chi_{c1}$)  and $\xi$-odd ($\Q=\chi_{c0,2}, h_c$) states.
It is easy to check that the parity under $\xi$ inversion is connected with the other quantum numbers of the particle by the relation
\begin{align}
  \label{eq:Pxi}
  P_\xi &=-C\sigma,
\end{align}
where $C$ is the charge parity of the particle.
Explicit parameterization of the distribution amplitudes can be found, for example in papers \cite{Braguta:2006wr, Braguta:2007fh, Braguta:2008qe, Ding:2015rkn, Hwang:2009cu} and we will discuss them in detail in section \ref{sec:numerical-results}.
 For further usage we collect all mentioned above quantum numbers in Table \ref{tab:quantum_numbers}.
If internal motion of quarks in meson is neglected (in the following we will refer to this limit as $\delta$-approximation) the distribution amplitudes $\phi_L^\Q(\xi)$ take the form
\begin{align}
  \label{eq:delta_approximation}
  \phiL{\eta_c,\JP,\chi_{c1}}(\xi) &= \delta(\xi),\quad \phiL{\chi_{c0,2},h_c}(\xi) = -\xi\delta'(\xi)
\end{align}
 of $\xi$-even and $\xi$-odd particles respectively. 

\begin{table}
  \centering
  \caption{Quantum numbers of charmonium mesons}
\begin{tabular}{@{}lllllr@{}}
\hline
 & $J^{PC}$ & $L$ & $S$ & $\sigma$ & $P_{\xi}$\\
\hline
$\eta_c$ & $0^{-+}$ & 0 & 0 & - & +\tabularnewline
$\JP$ & $1^{--}$ & 0 & 1 & + & +\tabularnewline
$\chi_{c0}$ & $0^{++}$ & 1 & 1 & + & -\tabularnewline
$\chi_{c1}$ & $1^{++}$ & 1 & 1 & - & +\tabularnewline
$\chi_{c2}$ & $2^{++}$ & 1 & 1 & + & -\tabularnewline
$h_c$ & $1^{+-}$ & 1 & 0 & - & -\tabularnewline
\hline
\end{tabular}
  \label{tab:quantum_numbers}
\end{table}

As it was noted above, the hard part of the matrix element can be calculated using perturbative QCD. At the leading order this element is described by shown in Fig.~\ref{fig:diags} diagrams. 
Corresponding to these diagrams matrix element takes the form
 \begin{align}
   \label{eq:LCmatr}
   \M(Z\to\Q_1\Q_2) &=\frac{64\pi\alpha_s}{9}c_{V,A} \frac{\fL{1}\fL{2}}{M_Z^2}
       I_{\Q_1\Q_2}\epsilon_Z^\mu (p_1-p_2)_\mu,
 \end{align}
where $p_{1,2}$ are the momenta of final charmonium, $\epsilon_Z$ is the initial $Z$-boson polarization vector, coupling constant $c_{V,A}$ is selected according to charge parity of the final state, and the integral $I_{\Q_1\Q_2}$ is defined as follows:
\begin{align}
  \label{eq:II}
  I_{\Q_1\Q_2} &= 
                 \frac{1}{2}\int_{-1}^1 d\xi_1 d\xi_2 \phiL{\Q_1}(\xi_1) \phiL{\Q_2}(\xi_2) 
                 \left[
                 \frac{1}{1+\xi_1}\frac{1}{1-\xi_2} - \frac{1}{1-\xi_1}\frac{1}{1+\xi_2}
                 \right].
\end{align}
The width of $Z\to\Q_1\Q_2$ decay is equal to
\begin{align}
 \label{eq:LCgamma}
  \Gamma(Z\to\Q_1\Q_2) &= \frac{256\pi\alpha_s^2}{243}c_{V,A}^2\frac{\fL{1}^2\fL{2}^2}{M_Z^3} I_{\Q_1\Q_2}^2
\end{align}
It is interesting to note, that coefficient \eqref{eq:II} is odd under the interchange $\xi_{1,2}\leftrightarrow -\xi_{1,2}$. Since, however, one of the final particles in the allowed processes \eqref{eq:allowed_proc} has $\xi$-even DA, while the other is $\xi$-odd, two terms in the integral do not cancel each other, but double the result.

In $\delta$-approximation \eqref{eq:delta_approximation} the integral in \eqref{eq:II} are actually equal to 1, so the width of the considered process takes a simple form
\begin{align}
  \label{eq:Gamma_delta}
  \Gamma(Z\to\Q_1\Q_2) &= \frac{256\pi\alpha_s^2}{243}c_{V,A}^2\frac{\fL{1}^2\fL{2}^2}{M_Z^3}  \end{align}
where, as it was discussed above, the constant $c_{V,A}$ is selected
according to charge parity of the final charmonium meson. It is easy to see that in $r\to0$ limit this relation coincide with NRQCD results \eqref{eq:NRQCD_widths_allowed1} -- \eqref{eq:NRQCD_widths_allowed2}.


\section{Numerical Results}
\label{sec:numerical-results}

Let us discuss numerical predictions for the branching fractions of the reactions under consideration and start with NRQCD results. The corresponding expressions for decay widths are presented in eq. \eqref{eq:NRQCD_widths_allowed1} -- \eqref{eq:NQRCD_widths_suppressed2}. Leptonic constants $\fN{\Q}$ are given in eq.~\eqref{eq:f_NRQCD_S}, \eqref{eq:f_NRQCD_P}, where NRQCD matrix elements $\langle O_1\rangle_{\JP,h_c}$ are equal to \cite{Braaten:2002fi}
\begin{align}
  \label{eq:NRQCD_O_num}
  \langle O_1\rangle_{\JP} &= 0.22\,\GeV^3,\qquad \langle O_1\rangle_{h_c} = 0.033\,\GeV^3.
\end{align}
These values correspond to
\begin{align}
  \label{eq:fNRQCD_num}
 \fN{\eta_c} &= \fN{\JP}=0.38,\GeV,\\
  \fN{\chi_{c0}} &= 0.057\,\GeV, \fN{\chi_{c1}}=0.14,\GeV, \\
  \fN{\chi_{c2}} &=0.081\,\GeV, \fN{h_c}=0.099\,\GeV.
\end{align}
The strong coupling constant $\alpha_s(\mu^2)$ is parametrized as
\begin{align}
  \label{eq:alphas}
  \alpha_s(\mu^2) &= \frac{4\pi}{b_0\ln(\mu^2/\Lambda_\mathrm{QCD}^2)}, \quad b_0=11-\frac{2}{3}n_f,
\end{align}
where $\Lambda_\mathrm{QCD}\approx 0.2\,\GeV$ and $n_f=5$ is the number of active flavors. At the scale $\mu^2=M_Z^2$ it corresponds to $\alpha_s(M_Z^2)\approx 0.13$. With presented above parameters it is easy to obtain the branching fractions of the decays, that are presented in the second column of table \ref{tab:tab}.

\begin{table}
  \centering
\begin{tabular}{cc|cccc}
\hline
 $\Q_1$ & $\Q_2$&  $\Br_\mathrm{NRQCD},\, 10^{-12}$ &  $\Br_\delta,\, 10^{-12}$ &  $\Br_\mathrm{LC},\, 10^{-12}$  & $\Br_{LC}/\Br_\delta$ \\ 
\hline 
$\eta_c$ & $ \eta_c$ &  --- & --- &  --- &  ---  \\ 
$\eta_c$ & $ J/\psi$ & $0.027$ &  --- &  --- &  ---  \\ 
$\eta_c$ & $ \chi_{c0}$ & $0.47$ &  $1.5\pm0.8_\mathrm{f}$ &  $2.3\pm1_\mathrm{f} \pm 0.2_\mathrm{wf}$ &  $1.6\pm0.1_\mathrm{wf}$ \\ 
$\eta_c$ & $ \chi_{c1}$ & $0.054$ &  --- &  --- &  ---  \\ 
$\eta_c$ & $ \chi_{c2}$ & $0.97$ &  $3.\pm2_\mathrm{f}$ &  $4.6\pm2_\mathrm{f} \pm 0.3_\mathrm{wf}$ &  $1.6\pm0.1_\mathrm{wf}$ \\ 
$\eta_c$ & $ h_c$ & $0.21$ &  $0.65\pm0.3_\mathrm{f}$ &  $1.\pm0.5_\mathrm{f} \pm 0.07_\mathrm{wf}$ &  $1.6\pm0.1_\mathrm{wf}$ \\ 
\hline
$J/\psi$ & $ J/\psi$ & $0.023$ &  --- &  --- &  ---  \\ 
$J/\psi$ & $ \chi_{c0}$ & $0.083$ &  $0.3\pm0.1_\mathrm{f}$ &  $0.47\pm0.2_\mathrm{f} \pm 0.03_\mathrm{wf}$ &  $1.6\pm0.1_\mathrm{wf}$ \\ 
$J/\psi$ & $ \chi_{c1}$ & $0.0035$ &  --- &  --- &  ---  \\ 
$J/\psi$ & $ \chi_{c2}$ & $0.14$ &  $0.6\pm0.3_\mathrm{f}$ &  $0.93\pm0.4_\mathrm{f} \pm 0.06_\mathrm{wf}$ &  $1.6\pm0.1_\mathrm{wf}$ \\ 
$J/\psi$ & $ h_c$ & $1.5$ &  $6.1\pm3_\mathrm{f}$ &  $9.5\pm5_\mathrm{f} \pm 0.6_\mathrm{wf}$ &  $1.6\pm0.1_\mathrm{wf}$ \\ 
\hline
$\chi_{c0}$ & $ \chi_{c0}$ &  --- & --- &  --- &  ---  \\ 
$\chi_{c0}$ & $ \chi_{c1}$ & $0.076$ &  $0.89\pm0.7_\mathrm{f}$ &  $1.4\pm1_\mathrm{f} \pm 0.09_\mathrm{wf}$ &  $1.5\pm0.1_\mathrm{wf}$ \\ 
$\chi_{c0}$ & $ \chi_{c2}$ & $0.0064$ &  --- &  --- &  ---  \\ 
$\chi_{c0}$ & $ h_c$ & $0.00035$ &  --- &  --- &  ---  \\ 
\hline
$\chi_{c1}$ & $ \chi_{c1}$ & $0.00039$ &  --- &  --- &  ---  \\ 
$\chi_{c1}$ & $ \chi_{c2}$ & $0.13$ &  $1.8\pm1_\mathrm{f}$ &  $2.8\pm2_\mathrm{f} \pm 0.2_\mathrm{wf}$ &  $1.5\pm0.1_\mathrm{wf}$ \\ 
$\chi_{c1}$ & $ h_c$ & $0.029$ &  $0.39\pm0.3_\mathrm{f}$ &  $0.61\pm0.5_\mathrm{f} \pm 0.04_\mathrm{wf}$ &  $1.5\pm0.1_\mathrm{wf}$ \\ 
\hline
$\chi_{c2}$ & $ \chi_{c2}$ & $0.00013$ &  --- &  --- &  ---  \\ 
$\chi_{c2}$ & $ h_c$ & $0.00023$ &  --- &  --- &  ---  \\ 
\hline
$h_c$ & $ h_c$ & $0.000099$ &  --- &  --- &  ---  \\ 
\hline 
\end{tabular}
\caption{Branching fractions of  $Z\to\Q_1\Q_2$ decays}
\label{tab:tab}
\end{table}

The branching fractions in the LC framework can be calculated using relation \eqref{eq:LCgamma}, but we need to know mesonic constants $\fL{\Q}$ and distribution amplitudes to use it. This question was studied in the literature thorougly, in the following we will use the results presented in \cite{Braguta:2006wr, Braguta:2007fh, Braguta:2008qe}. In particular, calculated on the scale $\mu=m_c$ leptonic constants of charmonia mesons are taken to be equal to
\begin{align}
  \label{eq:fLC_num}
  \fL{\eta_c}(m_c) &=(0.35\pm0.02)\GeV, \quad \fL{\JP}(m_c)=(0.41\pm0.02)\GeV,\quad
  \fL{\chi_{c0}}(m_c) = (0.11\pm0.02)\GeV,\\
 \fL{\chi_{c1}}(m_c) &= (0.27\pm0.05)\GeV,
  \fL{\chi_{c2}}(m_c)  = (0.16\pm0.03)\GeV, \quad \fL{h_c}(m_c) = (0.19\pm0.03)\GeV.
\end{align}
Note, the value of $c$ quark in LC model differs from phenomenological choice $M_{\JP}/2$ and is equal to $m_c=1.2\GeV$.

In the case of $\delta$-approximation one should use relations \eqref{eq:delta_approximation} for the distribution amplitudes. The corresponding results are shown in the third column of table \ref{tab:tab}, where marked with subscript ``$\mathrm{f}$'' errors are caused by uncertaines in the leptonic constants \eqref{eq:fLC_num}. One should note, the values of these constants differ a little bit from NRQCD case \eqref{eq:fNRQCD_num}. It should also be noted, that in LC framework outgoing mesons are assumed to be massless, so all mass corrections are neglected. For most of the processes this correction leads to uncertainty of about $2\%$. In the case of $Z\to\chi_{c0}\chi_{c1}$ decay, however the uncertainty caused by mass neglection is about $20\%$.


In order to calculate the branching fractions in full LC framework some parameterization for the functions $\phi_L^\Q(\xi)$ is required. Before we proceed to this point it is worth noting one important thing. The distribution amplitudes in eq.~\eqref{eq:LC_main} actually depend on the scale $\mu$, which is usually taken of the order of typical energy of the reaction. Presented in the literature results are given at the scale $\mu_0=m_c$ so we need to track the evolution of these functions to $\mu=M_Z$. In order do this it is convenient to write the function as a series over Gegenbauer polynomials
\begin{align}
  \label{eq:Gegenbauer}
  \phiL{\Q}(\xi,\mu_0) &= \sum_{n=n_0}^\infty a_n^\Q(\mu_0) C_n^{3/2}(\xi). 
\end{align}
It is clear, that for $\xi$-even and $\xi$-odd distribution amplitudes only even and odd coefficients are different from zero, so $n_0=0$ and $1$ in these two cases. According to \cite{Gribov:1972rt, Lipatov:1974qm, Altarelli:1977zs, Braun:2003rp} evolution results to the change of coefficients
\begin{align}
  a_n^\Q(\mu) &= L^{-\frac{\gamma_n}{b_0}} a_n^\Q(\mu_0),
\end{align}
where anomalous dimensions $\gamma_n$ for longitudinal current are equal to
\begin{align}
  \gamma_n = \frac{4}{3}\left(1-\frac{2}{(n+1)(n+2)}+\sum_{j=2}^{n+1}\frac{1}{j}\right)
\end{align}
and
\begin{align}
 L &= \frac{\alpha_s(\mu)}{\alpha_s(\mu_0)}.
\end{align}
As a result of this evolution the width of the distribution amplitude grows, but the normalization conditions \eqref{eq:norm_xi_even} are violated. If one wants to keep this normalization, it is possible to factor out the scale-dependent factor from the first term in the expansion \eqref{eq:Gegenbauer} and move it to the corresponding leptonic constant. As a result of this procedure the normalization of distribution amplitudes restores, but the evolution of the corresponding leptonic constant appears
\begin{align}
\phiL{\Q}(\xi,\mu) &= \sum_{n=n_0}^\infty 
                L^{-\frac{\gamma_n - \gamma_{n_0}}{b_0}}
a_n^\Q(\mu_0) C_n^{3/2}(\xi),
\quad
\fL{\Q}(\mu) = L^{-\frac{\gamma_{n_0}}{b_0}} \fL{\Q}(\mu_0).
\end{align}
 Since the first anomalous dimension $\gamma_0=0$, in the case of $\xi$-even states $\eta_c$, $J/\psi$, and $\chi_{c1}$ this evolution does change the constants. In the case of $\xi$-odd states $\chi_{c0,2}$, $h_c$, on the other hand with the increase of the scale leptonic constant decreases.

\begin{figure}
  \centering
  \includegraphics[width=0.9\textwidth]{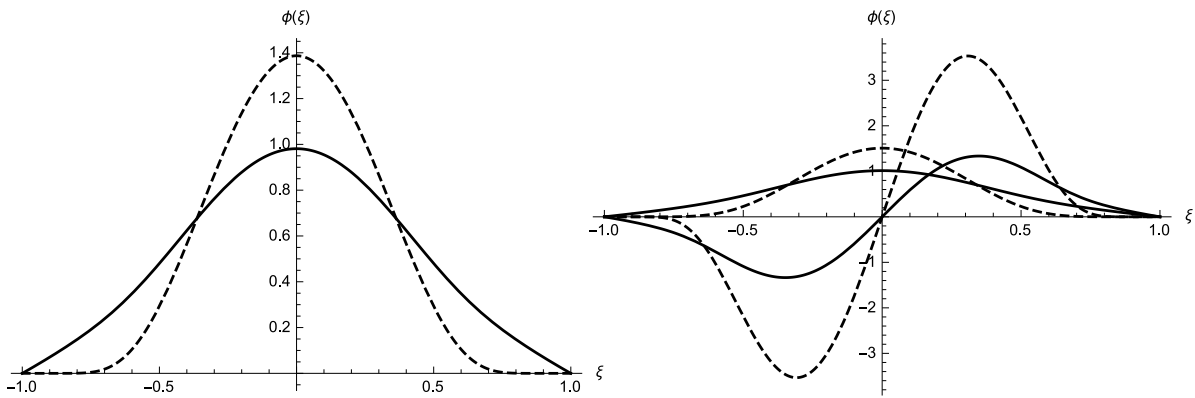}
  \caption{Distribution amplitudes of the $S$-wave (left figure) and $P$-wave (right figure) charmonium mesons. Dashed and solid lines correspond to $\mu^2=m_c^2$ and $\mu^2=M_Z^2$ scales respectively. In the case of $P$-wave states $\xi$-even curve correspond to $\chi_{c1}$ meson, while $\xi$-odd one to $\chi_{c0,2}$, $h_c$}
  \label{fig:wf}
\end{figure}

According to papers \cite{Braguta:2006wr, Braguta:2007fh} the distribution amplitude of $S$-wave charmonia can be written in the form
\begin{align}
  \phiL{\JP,\eta_c}(\xi,\mu_0) &= c(\beta_S) (1-\xi^2) \exp\left(-\frac{\beta_S}{1-\xi^2}\right),
\end{align}
where $c(\beta_S)$ is a normalization constant \eqref{eq:norm_xi_even} and
\begin{align}
  \beta_S &=3.8\pm0.7.
\end{align}
In the case of $P$-wave states there are two distribution functions \cite{Braguta:2008qe}:
\begin{align}
  \label{eq:6}
  \phiL{\chi_{c0,2},h_c}(\xi,\mu_0) &= c_1(\beta_P) \xi(1-\xi^2) \exp\left(-\frac{\beta_P}{1-\xi^2}\right),
\end{align}
for $\xi$-odd states and
\begin{align}
  \phiL{\chi_{c1}}(\xi,\mu_0) &= -c_2(\beta_P)\int_{-1}^\xi \phiL{h_c}(\xi,\mu_0)
\end{align}
for $\xi$-even. In these expressions $c_{1,2}(\beta_P)$ are  normalization constants \eqref{eq:norm_xi_even} and
\begin{align}
  \beta_P &=3.4^{+1.5}_{-0.9}.
\end{align}
In figure \ref{fig:wf} we show these distribution amplitudes at different scales. One can see that with the increase of the scale the effective widths of the DAs increase.

The results obtained using presented above distribution amplitudes are shown in the fourth column of table~\ref{tab:tab}. The errors labeled with subscript ``$\mathrm{wf}$'' in this column are caused by uncertainties in DAs parameters $\beta_{S,P}$. In order to show the effect of internal motion of the quarks in mesons in the fifth column of that table we present the ratio of branching fractions obtained in $\delta$-approximation and full LC. 
One can see, that in all allowed for leading twist  approximation processes internal quark motion leads to increase of the branching fraction.


\section{Conclusion}
\label{sec:conclusion}

The article is devoted to systematic analysis of double charmonium production in exclusive $Z$-boson decays. We present analytical and numerical results obtained using both Nonrelativistic Quantum Chromodynamics (NRQCD) model at color-singlet approximation and amplitude expansion on the light cone formalism. It turns out that helicity conservation rule restrict significantly the number of processes that can contribute at leading twist approximation and only longitudinally polarized mesons can be produced. Moreover, conservation of naturality quantum number sets even tighter restrictions and such processes as e.g. $Z\to 2\JP$ is forbidden completely at the leading twist approximation.

It is shown in the paper that NRQCD and LC results agree perfectly with each other and helicity suppression can ve clearly observed: branching fractions of the forbiden decays are suppressed by small chirality factor $r=m_c/M_Z$. Light-cone expansion formalism allows one to take into account internal motion of heavy quarks inside charmonium mesons and our calculations show that this effect results to increase of the widths by a factor about $1.5$. In spite if this increase theoretical predictions for the branching fractions are tiny (of the order of $10^{-12}$. We hope however, that considered processes can be observed experimentally, for example at LHC collider.

As it was noted above, in our analysis we restrict ourselves to color-singlet approximation. In the forthcoming paper we plan to analyze also contribution of color-octet components. It is also worth mentioning that the recent paper \cite{Ma:2017xno} a new Soft Gluon Factorization method (SGF) was proposed to describe charmonia production processes. In this approach velocity expansion is treated with more accuracy, so this model could be more suitable for calculations in comparison with NRQCD approach. In our future work we plan to study considered in our paper decays in SGF framework.


\begin{thebibliography}{24}%
\makeatletter
\providecommand \@ifxundefined [1]{%
 \@ifx{#1\undefined}
}%
\providecommand \@ifnum [1]{%
 \ifnum #1\expandafter \@firstoftwo
 \else \expandafter \@secondoftwo
 \fi
}%
\providecommand \@ifx [1]{%
 \ifx #1\expandafter \@firstoftwo
 \else \expandafter \@secondoftwo
 \fi
}%
\providecommand \natexlab [1]{#1}%
\providecommand \enquote  [1]{``#1''}%
\providecommand \bibnamefont  [1]{#1}%
\providecommand \bibfnamefont [1]{#1}%
\providecommand \citenamefont [1]{#1}%
\providecommand \href@noop [0]{\@secondoftwo}%
\providecommand \href [0]{\begingroup \@sanitize@url \@href}%
\providecommand \@href[1]{\@@startlink{#1}\@@href}%
\providecommand \@@href[1]{\endgroup#1\@@endlink}%
\providecommand \@sanitize@url [0]{\catcode `\\12\catcode `\$12\catcode
  `\&12\catcode `\#12\catcode `\^12\catcode `\_12\catcode `\%12\relax}%
\providecommand \@@startlink[1]{}%
\providecommand \@@endlink[0]{}%
\providecommand \url  [0]{\begingroup\@sanitize@url \@url }%
\providecommand \@url [1]{\endgroup\@href {#1}{\urlprefix }}%
\providecommand \urlprefix  [0]{URL }%
\providecommand \Eprint [0]{\href }%
\providecommand \doibase [0]{http://dx.doi.org/}%
\providecommand \selectlanguage [0]{\@gobble}%
\providecommand \bibinfo  [0]{\@secondoftwo}%
\providecommand \bibfield  [0]{\@secondoftwo}%
\providecommand \translation [1]{[#1]}%
\providecommand \BibitemOpen [0]{}%
\providecommand \bibitemStop [0]{}%
\providecommand \bibitemNoStop [0]{.\EOS\space}%
\providecommand \EOS [0]{\spacefactor3000\relax}%
\providecommand \BibitemShut  [1]{\csname bibitem#1\endcsname}%
\let\auto@bib@innerbib\@empty
\bibitem [{\citenamefont {Bodwin}\ \emph {et~al.}(1995)\citenamefont {Bodwin},
  \citenamefont {Braaten},\ and\ \citenamefont {Lepage}}]{Bodwin:1994jh}%
  \BibitemOpen
  \bibfield  {author} {\bibinfo {author} {\bibfnamefont {G.~T.}\ \bibnamefont
  {Bodwin}}, \bibinfo {author} {\bibfnamefont {E.}~\bibnamefont {Braaten}}, \
  and\ \bibinfo {author} {\bibfnamefont {G.~P.}\ \bibnamefont {Lepage}},\
  }\href {\doibase 10.1103/PhysRevD.55.5853, 10.1103/PhysRevD.51.1125}
  {\bibfield  {journal} {\bibinfo  {journal} {Phys. Rev.}\ }\textbf {\bibinfo
  {volume} {D51}},\ \bibinfo {pages} {1125} (\bibinfo {year} {1995})},\
  \bibinfo {note} {[Erratum: Phys. Rev.D55,5853(1997)]},\ \Eprint
  {http://arxiv.org/abs/hep-ph/9407339} {hep-ph/9407339} \BibitemShut {NoStop}%
\bibitem [{\citenamefont {Abe}\ \emph {et~al.}(2004)\citenamefont {Abe} \emph
  {et~al.}}]{Abe:2004ww}%
  \BibitemOpen
  \bibfield  {author} {\bibinfo {author} {\bibfnamefont {K.}~\bibnamefont
  {Abe}} \emph {et~al.} (\bibinfo {collaboration} {Belle}),\ }\href {\doibase
  10.1103/PhysRevD.70.071102} {\bibfield  {journal} {\bibinfo  {journal} {Phys.
  Rev.}\ }\textbf {\bibinfo {volume} {D70}},\ \bibinfo {pages} {071102}
  (\bibinfo {year} {2004})},\ \Eprint {http://arxiv.org/abs/hep-ex/0407009}
  {hep-ex/0407009} \BibitemShut {NoStop}%
\bibitem [{\citenamefont {Aubert}\ \emph {et~al.}(2005)\citenamefont {Aubert}
  \emph {et~al.}}]{Aubert:2005tj}%
  \BibitemOpen
  \bibfield  {author} {\bibinfo {author} {\bibfnamefont {B.}~\bibnamefont
  {Aubert}} \emph {et~al.} (\bibinfo {collaboration} {BaBar}),\ }\bibfield
  {booktitle} {\emph {\bibinfo {booktitle} {{Lepton and photon interactions at
  high energies. Proceedings, 22nd International Symposium, LP 2005, Uppsala,
  Sweden, June 30-July 5, 2005}}},\ }\href {\doibase
  10.1103/PhysRevD.72.031101} {\bibfield  {journal} {\bibinfo  {journal} {Phys.
  Rev.}\ }\textbf {\bibinfo {volume} {D72}},\ \bibinfo {pages} {031101}
  (\bibinfo {year} {2005})},\ \Eprint {http://arxiv.org/abs/hep-ex/0506062}
  {hep-ex/0506062} \BibitemShut {NoStop}%
\bibitem [{\citenamefont {Chernyak}\ and\ \citenamefont
  {Zhitnitsky}(1984)}]{Chernyak:1983ej}%
  \BibitemOpen
  \bibfield  {author} {\bibinfo {author} {\bibfnamefont {V.~L.}\ \bibnamefont
  {Chernyak}}\ and\ \bibinfo {author} {\bibfnamefont {A.~R.}\ \bibnamefont
  {Zhitnitsky}},\ }\href {\doibase 10.1016/0370-1573(84)90126-1} {\bibfield
  {journal} {\bibinfo  {journal} {Phys. Rept.}\ }\textbf {\bibinfo {volume}
  {112}},\ \bibinfo {pages} {173} (\bibinfo {year} {1984})}\BibitemShut
  {NoStop}%
\bibitem [{\citenamefont {Bondar}\ and\ \citenamefont
  {Chernyak}(2005)}]{Bondar:2004sv}%
  \BibitemOpen
  \bibfield  {author} {\bibinfo {author} {\bibfnamefont {A.~E.}\ \bibnamefont
  {Bondar}}\ and\ \bibinfo {author} {\bibfnamefont {V.~L.}\ \bibnamefont
  {Chernyak}},\ }\href {\doibase 10.1016/j.physletb.2005.03.021} {\bibfield
  {journal} {\bibinfo  {journal} {Phys. Lett.}\ }\textbf {\bibinfo {volume}
  {B612}},\ \bibinfo {pages} {215} (\bibinfo {year} {2005})},\ \Eprint
  {http://arxiv.org/abs/hep-ph/0412335} {hep-ph/0412335} \BibitemShut {NoStop}%
\bibitem [{\citenamefont {Braguta}\ \emph {et~al.}(2005)\citenamefont
  {Braguta}, \citenamefont {Likhoded},\ and\ \citenamefont
  {Luchinsky}}]{Braguta:2005kr}%
  \BibitemOpen
  \bibfield  {author} {\bibinfo {author} {\bibfnamefont {V.~V.}\ \bibnamefont
  {Braguta}}, \bibinfo {author} {\bibfnamefont {A.~K.}\ \bibnamefont
  {Likhoded}}, \ and\ \bibinfo {author} {\bibfnamefont {A.~V.}\ \bibnamefont
  {Luchinsky}},\ }\href {\doibase 10.1103/PhysRevD.72.074019} {\bibfield
  {journal} {\bibinfo  {journal} {Phys. Rev.}\ }\textbf {\bibinfo {volume}
  {D72}},\ \bibinfo {pages} {074019} (\bibinfo {year} {2005})},\ \Eprint
  {http://arxiv.org/abs/hep-ph/0507275} {hep-ph/0507275} \BibitemShut {NoStop}%
\bibitem [{\citenamefont {Braguta}\ \emph
  {et~al.}(2009{\natexlab{a}})\citenamefont {Braguta}, \citenamefont
  {Likhoded},\ and\ \citenamefont {Luchinsky}}]{Braguta:2009df}%
  \BibitemOpen
  \bibfield  {author} {\bibinfo {author} {\bibfnamefont {V.~V.}\ \bibnamefont
  {Braguta}}, \bibinfo {author} {\bibfnamefont {A.~K.}\ \bibnamefont
  {Likhoded}}, \ and\ \bibinfo {author} {\bibfnamefont {A.~V.}\ \bibnamefont
  {Luchinsky}},\ }\href {\doibase 10.1103/PhysRevD.85.119901,
  10.1103/PhysRevD.80.094008} {\bibfield  {journal} {\bibinfo  {journal} {Phys.
  Rev.}\ }\textbf {\bibinfo {volume} {D80}},\ \bibinfo {pages} {094008}
  (\bibinfo {year} {2009}{\natexlab{a}})},\ \bibinfo {note} {[Erratum: Phys.
  Rev.D85,119901(2012)]},\ \Eprint {http://arxiv.org/abs/0902.0459}
  {arXiv:0902.0459 [hep-ph]} \BibitemShut {NoStop}%
\bibitem [{\citenamefont {Braguta}\ and\ \citenamefont
  {Kartvelishvili}(2010)}]{Braguta:2009xu}%
  \BibitemOpen
  \bibfield  {author} {\bibinfo {author} {\bibfnamefont {V.~V.}\ \bibnamefont
  {Braguta}}\ and\ \bibinfo {author} {\bibfnamefont {V.~G.}\ \bibnamefont
  {Kartvelishvili}},\ }\href {\doibase 10.1103/PhysRevD.81.014012} {\bibfield
  {journal} {\bibinfo  {journal} {Phys. Rev.}\ }\textbf {\bibinfo {volume}
  {D81}},\ \bibinfo {pages} {014012} (\bibinfo {year} {2010})},\ \Eprint
  {http://arxiv.org/abs/0907.2772} {arXiv:0907.2772 [hep-ph]} \BibitemShut
  {NoStop}%
\bibitem [{\citenamefont {Chen}\ and\ \citenamefont
  {Qiao}(2012)}]{Chen:2012ih}%
  \BibitemOpen
  \bibfield  {author} {\bibinfo {author} {\bibfnamefont {L.-B.}\ \bibnamefont
  {Chen}}\ and\ \bibinfo {author} {\bibfnamefont {C.-F.}\ \bibnamefont
  {Qiao}},\ }\href {\doibase 10.1007/JHEP11(2012)168} {\bibfield  {journal}
  {\bibinfo  {journal} {JHEP}\ }\textbf {\bibinfo {volume} {11}},\ \bibinfo
  {pages} {168} (\bibinfo {year} {2012})},\ \Eprint
  {http://arxiv.org/abs/1204.0215} {arXiv:1204.0215 [hep-ph]} \BibitemShut
  {NoStop}%
\bibitem [{\citenamefont {Guberina}\ \emph {et~al.}(1980)\citenamefont
  {Guberina}, \citenamefont {Kuhn}, \citenamefont {Peccei},\ and\ \citenamefont
  {Ruckl}}]{Guberina:1980dc}%
  \BibitemOpen
  \bibfield  {author} {\bibinfo {author} {\bibfnamefont {B.}~\bibnamefont
  {Guberina}}, \bibinfo {author} {\bibfnamefont {J.~H.}\ \bibnamefont {Kuhn}},
  \bibinfo {author} {\bibfnamefont {R.~D.}\ \bibnamefont {Peccei}}, \ and\
  \bibinfo {author} {\bibfnamefont {R.}~\bibnamefont {Ruckl}},\ }\href
  {\doibase 10.1016/0550-3213(80)90287-4} {\bibfield  {journal} {\bibinfo
  {journal} {Nucl. Phys.}\ }\textbf {\bibinfo {volume} {B174}},\ \bibinfo
  {pages} {317} (\bibinfo {year} {1980})}\BibitemShut {NoStop}%
\bibitem [{\citenamefont {Grossman}\ \emph {et~al.}(2015)\citenamefont
  {Grossman}, \citenamefont {KÃ¶nig},\ and\ \citenamefont
  {Neubert}}]{Grossmann:2015lea}%
  \BibitemOpen
  \bibfield  {author} {\bibinfo {author} {\bibfnamefont {Y.}~\bibnamefont
  {Grossman}}, \bibinfo {author} {\bibfnamefont {M.}~\bibnamefont {KÃ¶nig}}, \
  and\ \bibinfo {author} {\bibfnamefont {M.}~\bibnamefont {Neubert}},\ }\href
  {\doibase 10.1007/JHEP04(2015)101} {\bibfield  {journal} {\bibinfo  {journal}
  {JHEP}\ }\textbf {\bibinfo {volume} {04}},\ \bibinfo {pages} {101} (\bibinfo
  {year} {2015})},\ \Eprint {http://arxiv.org/abs/1501.06569} {arXiv:1501.06569
  [hep-ph]} \BibitemShut {NoStop}%
\bibitem [{\citenamefont {Alte}\ \emph {et~al.}(2016)\citenamefont {Alte},
  \citenamefont {Grossman}, \citenamefont {KÃ¶nig},\ and\ \citenamefont
  {Neubert}}]{Alte:2017ycm}%
  \BibitemOpen
  \bibfield  {author} {\bibinfo {author} {\bibfnamefont {S.}~\bibnamefont
  {Alte}}, \bibinfo {author} {\bibfnamefont {Y.}~\bibnamefont {Grossman}},
  \bibinfo {author} {\bibfnamefont {M.}~\bibnamefont {KÃ¶nig}}, \ and\ \bibinfo
  {author} {\bibfnamefont {M.}~\bibnamefont {Neubert}},\ }\bibfield
  {booktitle} {\emph {\bibinfo {booktitle} {{Proceedings, 38th International
  Conference on High Energy Physics (ICHEP 2016): Chicago, IL, USA, August
  3-10, 2016}}},\ }\href@noop {} {\bibfield  {journal} {\bibinfo  {journal}
  {PoS}\ }\textbf {\bibinfo {volume} {ICHEP2016}},\ \bibinfo {pages} {618}
  (\bibinfo {year} {2016})},\ \Eprint {http://arxiv.org/abs/1703.07242}
  {arXiv:1703.07242 [hep-ph]} \BibitemShut {NoStop}%
\bibitem [{\citenamefont {Luchinsky}(2017)}]{Luchinsky:2017jab}%
  \BibitemOpen
  \bibfield  {author} {\bibinfo {author} {\bibfnamefont {A.~V.}\ \bibnamefont
  {Luchinsky}},\ }\href@noop {} {\  (\bibinfo {year} {2017})},\ \Eprint
  {http://arxiv.org/abs/1706.04091} {arXiv:1706.04091 [hep-ph]} \BibitemShut
  {NoStop}%
\bibitem [{\citenamefont {Braaten}\ and\ \citenamefont
  {Lee}(2003)}]{Braaten:2002fi}%
  \BibitemOpen
  \bibfield  {author} {\bibinfo {author} {\bibfnamefont {E.}~\bibnamefont
  {Braaten}}\ and\ \bibinfo {author} {\bibfnamefont {J.}~\bibnamefont {Lee}},\
  }\href {\doibase 10.1103/PhysRevD.72.099901, 10.1103/PhysRevD.67.054007}
  {\bibfield  {journal} {\bibinfo  {journal} {Phys. Rev.}\ }\textbf {\bibinfo
  {volume} {D67}},\ \bibinfo {pages} {054007} (\bibinfo {year} {2003})},\
  \bibinfo {note} {[Erratum: Phys. Rev.D72,099901(2005)]},\ \Eprint
  {http://arxiv.org/abs/hep-ph/0211085} {hep-ph/0211085} \BibitemShut {NoStop}%
\bibitem [{\citenamefont {Braguta}\ \emph {et~al.}(2007)\citenamefont
  {Braguta}, \citenamefont {Likhoded},\ and\ \citenamefont
  {Luchinsky}}]{Braguta:2006wr}%
  \BibitemOpen
  \bibfield  {author} {\bibinfo {author} {\bibfnamefont {V.~V.}\ \bibnamefont
  {Braguta}}, \bibinfo {author} {\bibfnamefont {A.~K.}\ \bibnamefont
  {Likhoded}}, \ and\ \bibinfo {author} {\bibfnamefont {A.~V.}\ \bibnamefont
  {Luchinsky}},\ }\href {\doibase 10.1016/j.physletb.2007.01.014} {\bibfield
  {journal} {\bibinfo  {journal} {Phys. Lett.}\ }\textbf {\bibinfo {volume}
  {B646}},\ \bibinfo {pages} {80} (\bibinfo {year} {2007})},\ \Eprint
  {http://arxiv.org/abs/hep-ph/0611021} {hep-ph/0611021} \BibitemShut {NoStop}%
\bibitem [{\citenamefont {Braguta}(2007)}]{Braguta:2007fh}%
  \BibitemOpen
  \bibfield  {author} {\bibinfo {author} {\bibfnamefont {V.~V.}\ \bibnamefont
  {Braguta}},\ }\href {\doibase 10.1103/PhysRevD.75.094016} {\bibfield
  {journal} {\bibinfo  {journal} {Phys. Rev.}\ }\textbf {\bibinfo {volume}
  {D75}},\ \bibinfo {pages} {094016} (\bibinfo {year} {2007})},\ \Eprint
  {http://arxiv.org/abs/hep-ph/0701234} {hep-ph/0701234} \BibitemShut {NoStop}%
\bibitem [{\citenamefont {Braguta}\ \emph
  {et~al.}(2009{\natexlab{b}})\citenamefont {Braguta}, \citenamefont
  {Likhoded},\ and\ \citenamefont {Luchinsky}}]{Braguta:2008qe}%
  \BibitemOpen
  \bibfield  {author} {\bibinfo {author} {\bibfnamefont {V.~V.}\ \bibnamefont
  {Braguta}}, \bibinfo {author} {\bibfnamefont {A.~K.}\ \bibnamefont
  {Likhoded}}, \ and\ \bibinfo {author} {\bibfnamefont {A.~V.}\ \bibnamefont
  {Luchinsky}},\ }\href {\doibase 10.1103/PhysRevD.79.074004} {\bibfield
  {journal} {\bibinfo  {journal} {Phys. Rev.}\ }\textbf {\bibinfo {volume}
  {D79}},\ \bibinfo {pages} {074004} (\bibinfo {year} {2009}{\natexlab{b}})},\
  \Eprint {http://arxiv.org/abs/0810.3607} {arXiv:0810.3607 [hep-ph]}
  \BibitemShut {NoStop}%
\bibitem [{\citenamefont {Ding}\ \emph {et~al.}(2016)\citenamefont {Ding},
  \citenamefont {Gao}, \citenamefont {Chang}, \citenamefont {Liu},\ and\
  \citenamefont {Roberts}}]{Ding:2015rkn}%
  \BibitemOpen
  \bibfield  {author} {\bibinfo {author} {\bibfnamefont {M.}~\bibnamefont
  {Ding}}, \bibinfo {author} {\bibfnamefont {F.}~\bibnamefont {Gao}}, \bibinfo
  {author} {\bibfnamefont {L.}~\bibnamefont {Chang}}, \bibinfo {author}
  {\bibfnamefont {Y.-X.}\ \bibnamefont {Liu}}, \ and\ \bibinfo {author}
  {\bibfnamefont {C.~D.}\ \bibnamefont {Roberts}},\ }\href {\doibase
  10.1016/j.physletb.2015.11.075} {\bibfield  {journal} {\bibinfo  {journal}
  {Phys. Lett.}\ }\textbf {\bibinfo {volume} {B753}},\ \bibinfo {pages} {330}
  (\bibinfo {year} {2016})},\ \Eprint {http://arxiv.org/abs/1511.04943}
  {arXiv:1511.04943 [nucl-th]} \BibitemShut {NoStop}%
\bibitem [{\citenamefont {Hwang}(2009)}]{Hwang:2009cu}%
  \BibitemOpen
  \bibfield  {author} {\bibinfo {author} {\bibfnamefont {C.-W.}\ \bibnamefont
  {Hwang}},\ }\href {\doibase 10.1088/1126-6708/2009/10/074} {\bibfield
  {journal} {\bibinfo  {journal} {JHEP}\ }\textbf {\bibinfo {volume} {10}},\
  \bibinfo {pages} {074} (\bibinfo {year} {2009})},\ \Eprint
  {http://arxiv.org/abs/0906.4412} {arXiv:0906.4412 [hep-ph]} \BibitemShut
  {NoStop}%
\bibitem [{\citenamefont {Gribov}\ and\ \citenamefont
  {Lipatov}(1972)}]{Gribov:1972rt}%
  \BibitemOpen
  \bibfield  {author} {\bibinfo {author} {\bibfnamefont {V.~N.}\ \bibnamefont
  {Gribov}}\ and\ \bibinfo {author} {\bibfnamefont {L.~N.}\ \bibnamefont
  {Lipatov}},\ }\href@noop {} {\bibfield  {journal} {\bibinfo  {journal} {Sov.
  J. Nucl. Phys.}\ }\textbf {\bibinfo {volume} {15}},\ \bibinfo {pages} {675}
  (\bibinfo {year} {1972})},\ \bibinfo {note} {[Yad.
  Fiz.15,1218(1972)]}\BibitemShut {NoStop}%
\bibitem [{\citenamefont {Lipatov}(1975)}]{Lipatov:1974qm}%
  \BibitemOpen
  \bibfield  {author} {\bibinfo {author} {\bibfnamefont {L.~N.}\ \bibnamefont
  {Lipatov}},\ }\href@noop {} {\bibfield  {journal} {\bibinfo  {journal} {Sov.
  J. Nucl. Phys.}\ }\textbf {\bibinfo {volume} {20}},\ \bibinfo {pages} {94}
  (\bibinfo {year} {1975})},\ \bibinfo {note} {[Yad.
  Fiz.20,181(1974)]}\BibitemShut {NoStop}%
\bibitem [{\citenamefont {Altarelli}\ and\ \citenamefont
  {Parisi}(1977)}]{Altarelli:1977zs}%
  \BibitemOpen
  \bibfield  {author} {\bibinfo {author} {\bibfnamefont {G.}~\bibnamefont
  {Altarelli}}\ and\ \bibinfo {author} {\bibfnamefont {G.}~\bibnamefont
  {Parisi}},\ }\href {\doibase 10.1016/0550-3213(77)90384-4} {\bibfield
  {journal} {\bibinfo  {journal} {Nucl. Phys.}\ }\textbf {\bibinfo {volume}
  {B126}},\ \bibinfo {pages} {298} (\bibinfo {year} {1977})}\BibitemShut
  {NoStop}%
\bibitem [{\citenamefont {Braun}\ \emph {et~al.}(2003)\citenamefont {Braun},
  \citenamefont {Korchemsky},\ and\ \citenamefont {Mueller}}]{Braun:2003rp}%
  \BibitemOpen
  \bibfield  {author} {\bibinfo {author} {\bibfnamefont {V.~M.}\ \bibnamefont
  {Braun}}, \bibinfo {author} {\bibfnamefont {G.~P.}\ \bibnamefont
  {Korchemsky}}, \ and\ \bibinfo {author} {\bibfnamefont {D.}~\bibnamefont
  {Mueller}},\ }\href {\doibase 10.1016/S0146-6410(03)90004-4} {\bibfield
  {journal} {\bibinfo  {journal} {Prog. Part. Nucl. Phys.}\ }\textbf {\bibinfo
  {volume} {51}},\ \bibinfo {pages} {311} (\bibinfo {year} {2003})},\ \Eprint
  {http://arxiv.org/abs/hep-ph/0306057} {hep-ph/0306057} \BibitemShut {NoStop}%
\bibitem [{\citenamefont {Ma}\ and\ \citenamefont {Chao}(2017)}]{Ma:2017xno}%
  \BibitemOpen
  \bibfield  {author} {\bibinfo {author} {\bibfnamefont {Y.-Q.}\ \bibnamefont
  {Ma}}\ and\ \bibinfo {author} {\bibfnamefont {K.-T.}\ \bibnamefont {Chao}},\
  }\href@noop {} {\  (\bibinfo {year} {2017})},\ \Eprint
  {http://arxiv.org/abs/1703.08402} {arXiv:1703.08402 [hep-ph]} \BibitemShut
  {NoStop}%
\end{thebibliography}

%

\appendix

\section{NRQCD Matrix Elements}
\label{sec:nrqcd-matr-elem}

Squared matrix element of double charmonia production in exclusive $Z$-bozon decays can be written as
\begin{align}
  \left|\M(Z(\lambda_Z)\to\Q_1(\lambda_1)\Q_2(\lambda_2)  \right|^2 &=
   \frac{4096\pi^2\alpha_s^2}{81}
    c_{V,A}^2
    \frac{\fN{\Q_1}^2\fN{\Q_2}^2}{M_Z^2} C^{\Q_1\Q_2}_{\lambda_1\lambda_2},
\end{align}
where $c_{V,A}$ are vector and axial coupling constants of $Zc\bar{c}$ vertex \eqref{eq:Zvert} and $\fN{\Q}$ are NRQCD mesonic constants defined in \eqref{eq:f_NRQCD_S}. 
From orbital momentum conservation it follows the $\lambda_Z=\lambda_1-\lambda_2$. It is also evident that
\begin{align}
  C^{\Q_1\Q_2}_{-\lambda_1-\lambda_2} &= C^{\Q_1\Q_2}_{\lambda_1\lambda_2}.
\end{align}
Nonzero values of these coefficients are equal to
\begin{align}
C^{\eta_c J/\psi}_{01} &=4 \beta ^2 r^2,\quad
\\ 
C^{\eta_c \chi_{c0}}_{00} &=\beta ^2 \left(1-2 r^2\right)^2,\quad
\\ 
C^{\eta_c \chi_{c1}}_{01} &=9 \beta ^2 r^2 \left(r^2-1\right)^2,\quad
\\ 
C^{\eta_c \chi_{c2}}_{00} &=\beta ^2 \left(r^2+1\right)^2,\quad
C^{\eta_c \chi_{c2}}_{01} =3 \beta ^2 r^2 \left(2-3 r^2\right)^2,\quad
\\ 
C^{\eta_c h_c}_{00} &=\left(12 r^4-6 r^2+1\right)^2,\quad
C^{\eta_c h_c}_{01} =r^2 \left(1-2 r^2\right)^2,\quad
\\ 
C^{J/\psi J/\psi}_{01} &=\beta ^4 r^2,\quad
\\ 
C^{J/\psi \chi_{c0}}_{00} &=\left(-12 r^4+10 r^2+1\right)^2,\quad
C^{J/\psi \chi_{c0}}_{10} =r^2 \left(9-14 r^2\right)^2,\quad
\\ 
C^{J/\psi \chi_{c1}}_{01} &=r^2 \left(2-7 r^2\right)^2,\quad
C^{J/\psi \chi_{c1}}_{10} =r^6,\quad
C^{J/\psi \chi_{c1}}_{11} =4 r^4 \left(1-3 r^2\right)^2,\quad
\\ 
C^{J/\psi \chi_{c2}}_{00} &=\left(12 r^4+2 r^2-1\right)^2,\quad
C^{J/\psi \chi_{c2}}_{01} =3 r^2 \left(1-5 r^2\right)^2,\quad
C^{J/\psi \chi_{c2}}_{10} =r^2 \left(3-11 r^2\right)^2,
\\
C^{J/\psi \chi_{c2}}_{11} &=12 r^4 \left(1-3 r^2\right)^2,\quad
C^{J/\psi \chi_{c2}}_{12} =6 r^6,\quad
\\ 
C^{J/\psi h_c}_{00} &=\beta ^2,\quad
C^{J/\psi h_c}_{10} =4 \beta ^2 r^2 \left(2-3 r^2\right)^2,\quad
C^{J/\psi h_c}_{11} =\beta ^2 r^4,\quad
\\ 
C^{\chi_{c0} \chi_{c1}}_{00} &=\left(-4 r^4+5 r^2+1\right)^2,\quad
C^{\chi_{c0} \chi_{c1}}_{01} =4 r^2 \left(8 r^4-14 r^2+5\right)^2,\quad
\\ 
C^{\chi_{c0} \chi_{c2}}_{01} &=3 r^2 \left(24 r^4-34 r^2+7\right)^2,\quad
\\ 
C^{\chi_{c0} h_c}_{01} &=4 \beta ^2 r^2 \left(3-2 r^2\right)^2,\quad
\\ 
C^{\chi_{c1} \chi_{c1}}_{01} &=\frac{1}{4} r^2 \left(4 r^4-9 r^2+2\right)^2,\quad
\\ 
C^{\chi_{c1} \chi_{c2}}_{00} &=\left(4 r^4-\beta ^2\right)^2,\quad
C^{\chi_{c1} \chi_{c2}}_{01} =\frac{3}{4} r^6 \left(4 r^2+1\right)^2,\quad
C^{\chi_{c1} \chi_{c2}}_{10} =\frac{1}{4} r^2 \left(44 r^4-41 r^2+8\right)^2,\\
C^{\chi_{c1} \chi_{c2}}_{11} &=3 r^4 \left(1-2 r^2\right)^2,\quad
C^{\chi_{c1} \chi_{c2}}_{12} =\frac{3}{2} r^6 \left(3-4 r^2\right)^2,\quad
\\ 
C^{\chi_{c1} h_c}_{00} &=\beta ^2,\quad
C^{\chi_{c1} h_c}_{01} =\beta ^2 r^2 \left(1-2 r^2\right)^2,\quad
C^{\chi_{c1} h_c}_{10} =25 \beta ^2 r^2 \left(1-2 r^2\right)^2,\quad
\\ 
C^{\chi_{c2} \chi_{c2}}_{01} &=\frac{3}{4} r^2 \left(36 r^4-17 r^2+2\right)^2,\quad
C^{\chi_{c2} \chi_{c2}}_{12} =\frac{9 \beta ^4 r^6}{2},\quad
\\ 
C^{\chi_{c2} h_c}_{01} &=4 \beta ^2 r^6,\quad
C^{\chi_{c2} h_c}_{10} =12 \beta ^2 r^2 \left(1-5 r^2\right)^2,\quad
C^{\chi_{c2} h_c}_{21} =24 \beta ^2 r^6,\quad
\\ 
C^{h_c h_c}_{01} &=\beta ^4 r^2
\end{align}
where
\begin{align}
  r &=\frac{m_c}{M_Z},\qquad \beta=\sqrt{1-16r^2}.
\end{align}
One can easily see, that in agreement with discussed earlier selection rules, production of transversly polarized particles are suppressed by the chirality factor $\sim r^{2|\lambda_1+\lambda_2|}$. In the case of longitdial meson production, on the other hand, in $r\to0$ limit we have $C^{\Q_1\Q_2}_{00}=1$.

\end{document}